\documentclass[12pt]{article}
 \def\gae{\; ^{>}_{\sim} \;}

\title{\textbf{REDUCING THE SPECTRAL INDEX IN SUPERNATURAL INFLATION}}
{\author{\\[1cm]
{\sc \large Chia-Min Lin$^{1}$ and Kingman Cheung$^{2}$}\\
{\sl\small Department of Physics, National Tsing Hua University, Hsinchu, Taiwan 300 }\\
{\sl\small Physics Division, National Center for Theoretical Sciences,
Hsinchu 300, Taiwan}
}}

\usepackage[margin=2cm]{geometry}
\usepackage{graphicx,color}
\usepackage{subfigure}
\begin{document}
\maketitle
\begin{abstract}
  Supernatural inflation is an attractive model based just on a flat
  direction with soft SUSY breaking mass terms in the framework of
  supersymmetry. The beauty of the model is inferred from its name
  that the model needs no fine-tuning. However, the prediction of the
  spectral index is $n_s \gae 1$, in contrast to
  experimental data. In this paper, we show that the beauty of
  supernatural inflation with the spectral index reduced to
  $n_s=0.96$ without any fine-tuning, by considering the general
  feature that a flat direction is lifted by a non-renormalizable term
  with an A-term.
\end{abstract}
\footnoterule{\small $^1$cmlin@phys.nthu.edu.tw, $^2$cheung@phys.nthu.edu.tw}
\section{Introduction}
Inflation \cite{Starobinsky:1980te, Sato:1980yn, Guth:1980zm} (for
review, \cite{Lyth:1998xn, Lyth:2007qh, Linde:2007fr}) is an vacuum-dominated
epoch in the early Universe when the scale factor grew
exponentially. This scenario is used to set the initial condition for
the hot big bang model and to provide the primordial density perturbation as the
seed of structure formation. In the framework of slow-roll
inflation, the slow-roll parameters are defined by
\begin{equation}
\epsilon \equiv \frac{M_P^2}{2} \left(\frac{V^\prime}{V}\right)^2,
\end{equation}
\begin{equation}
\eta \equiv M_P^2\frac{V^{\prime\prime}}{V},
\end{equation}
where $M_P=2.4\times 10^{18} \mbox{ GeV}$ is the reduced Planck mass.
The spectral index can be expressed in terms of the slow-roll parameters as
\begin{equation}
n_s=1+2\eta-6\epsilon.
\label{n}
\end{equation}
The latest WMAP 5-year result prefers the spectral index around
$n_s=0.96$ \cite{Komatsu:2008hk}.
The spectrum is given by
\begin{equation}
P_R=\frac{1}{12\pi^2M_P^6}\frac{V^3}{V'^2} \;.
\end{equation}
With the slow-roll approximation
the value of the inflaton field $\phi$,
in order to achieve $N$ e-folds inflation, is
\begin{equation}
N=M^{-2}_P\int^{\phi(N)}_{\phi_{end}}\frac{V}{V'}d\phi.
\label{efolds}
\end{equation}
From observation \cite{Komatsu:2008hk} $P^{1/2}_R\simeq 5\times 10^{-5}$ at $N \simeq 60$ (we call this CMB normalization).

In order to build a successful inflation model, we need
a potential which is very flat when $N \simeq 60$ and becomes steep
when inflation ends. It is difficult to
achieve this form of potential
by a single field without significant tuning of the coupling
parameter \cite{Lyth:1998xn}.  The idea of hybrid inflation is
more natural in achieve so, in which
the jobs of ending the inflation and providing the scale of inflation is
done by another scalar field (called waterfall field).   The quadratic
potential for the inflaton field $\phi$ is
\begin{equation}
V=\frac{1}{2}m^2\phi^2.
\end{equation}
This potential is used in the case of chaotic inflation
\cite{Linde:1983gd} where $\phi>M_P$ is required. It can be turned
into a hybrid inflation \cite{Linde:1993cn} by adding a ``false
vacuum'' to it \cite{Copeland:1994vg}:
\begin{equation}
V=V_0+\frac{1}{2}m^2\phi^2.
\end{equation}
Consequently we can have an inflation model with $\phi<M_P$.  In
this case, the end of inflation is not due to the failing of
slow-roll, but the tachyonic instability of the waterfall
field.

The idea of supernatural inflation \cite{Randall:1995dj} is that the
inflaton field $\phi$ is a flat direction
in supersymmetric field theory and the mass term is provided by
a soft SUSY breaking term and $V_0$ by another
field coupled to the inflaton field.  Therefore, the model is basically
a tree-level hybrid inflation model. However, this model predicts a
spectral index $n_s \gae 1$. A method to reduce the spectral index in
this tree-level hybrid inflation is by converting the model into ``hilltop
inflation'' \cite{Boubekeur:2005zm} via introduction of a negative quartic
term to the potential as shown in \cite{Kohri:2007gq}. The only
problem here is how to introduce this quartic term and whether the
coupling parameter needs fine tuning. The answer relies on the fact
that the flat direction is generically
affected not only by the soft
SUSY breaking term but also the nonrenormalizable term and A-term. In
this paper, we show that by considering these additional terms the spectral
index can be reduced to $n_s=0.96$ in a natural fashion.

This paper is organized as follows. In Sec.~\ref{2}, we introduce
the scalar potential of the model. In Sec.~\ref{3}, the analytic
solution base on the potential and the main result of this paper are
provided. Finally, we present our conclusions in Sec.~\ref{5}.

\section{The Potential}
\label{2}
Suppose we want to build an inflation model based on a SUSY flat
direction \cite{Gherghetta:1995dv}, we have to know that generically
a flat direction is lifted by supersymmetry breaking terms and
non-renormalizable superpotential terms \cite{Dine:1995uk,Dine:1995kz},
which has the form
\begin{equation}
W=\lambda_p \frac{\phi^p}{M^{p-3}_P}
\end{equation}
where $\phi$ is the flat direction and $4 \leq p \leq 9$
\cite{Gherghetta:1995dv}.  Therefore, the potential is
\cite{Allahverdi:2006iq, Allahverdi:2006we, Bueno Sanchez:2006xk, Lyth:2006ec} (after minimizing the
potential along the angular direction)
\begin{equation}
V=\frac{1}{2}m^2\phi^2-A\frac{\lambda_p \phi^p}{p M^{p-3}_P}
+\lambda^2_p\frac{\phi^{2(p-1)}}{M_P^{2(p-3)}}.
\end{equation}
By spirit of hybrid inflation we can add a ``false vacuum'' to this
potential via a coupling to a waterfall field similar to the case of
supernatural inflation. Therefore, the potential we consider is of the
following form
\begin{equation}
V(\phi)=V_0+\frac{1}{2}m^2\phi^2-\frac{\lambda_p A \phi^p}{p M_P}+
\lambda_p^2\frac{\phi^{2(p-1)}}{M_P^{2(p-3)}}
\label{potential2}
\end{equation}
If we just consider the first two terms, the result is the tree-level
hybrid inflation, which is realized in supernatural inflation where the
mass term comes from soft SUSY breaking.  In this work, we
focus on the case of $p=4$ (the least of Planck-mass suppression),
and neglect the last term (which is
possible when $\phi \ll M_P$ and will be justified in the following
section). Therefore, the potential is
\begin{eqnarray}
V(\phi)&=&V_0+\frac{1}{2}m^2\phi^2-\frac{\lambda_4 A \phi^4}{4M_P}\\
       &\equiv& V_0 \left(1+\frac{1}{2}\eta_0\frac{\phi^2}{M_P^2}\right)-\lambda \phi^4
\label{potential}
\end{eqnarray}
with
\begin{equation}
\eta_0 \equiv \frac{m^2 M_P^2}{V_0}  \;\;\; \mbox{and} \;\;\; \lambda \equiv \frac{\lambda_4 A}{4 M_P}
\end{equation}
This form of potential has been considered in \cite{Kohri:2007gq}.
The question is whether $\eta_0$ and $\lambda$ here need
fine-tuning. The natural value of soft SUSY breaking terms,
$m$ and $A$,  are $m \sim A \sim
O(\mbox{TeV}) \sim 10^{-15} M_P$. The coupling $\lambda_4$ is of
$O(1)$, which makes
$\lambda \sim O(10^{-15})$. As in the case of supernatural
inflation, we choose $V_0=M_I^4$ where $M_I \simeq 10^{11} \mbox{GeV}
\simeq 10^{-7} M_P$ is the intermediate scale, therefore
$\eta_0=O(10^{-2})$. In the following section, we will apply those
natural values without fine-tuning to achieve our goal of reducing the
spectral index of supernatural inflation.

\section{Analytical Solution}
\label{3}
\begin{figure}[htbp]
 \begin{center}
 \includegraphics[width=0.45\textwidth]{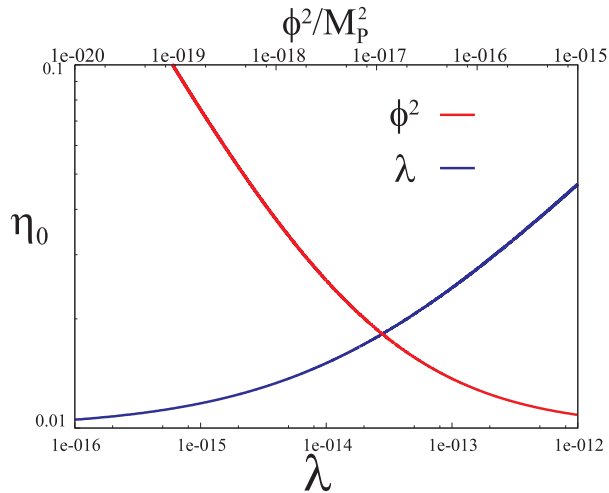}
 \caption{$\phi^2/M_P^2$ and $\lambda$ as a function of $\eta_0$ at $N=60$.}
 \label{result}
 \end{center}
 \end{figure}

From Eq. (\ref{potential}), by using Eq. (\ref{n}-\ref{efolds}) we can obtain
\begin{eqnarray}
\left(\frac{\phi}{M_P}\right)^2&=&\left(\frac{V_0}{M_P^4}\right)
\frac{\eta_0 e^{2N\eta_0}}{\eta_0 x+4 \lambda (e^{2N\eta_0}-1)}\\
x &
\equiv & \left(\frac{V_0}{M_P^4}\right) \left(\frac{M_P}{\phi_{end}}\right)^2,
\end{eqnarray}
and
\begin{eqnarray}
P_R&=&\frac{1}{12\pi^2}e^{-2N\eta_0}\frac{[4\lambda(e^{2N\eta_0}-1)+\eta_0 x]^3}{\eta_0^3(\eta_0 x-4\lambda)^2}\\
n_s&=&1+2\eta_0 \left[1-\frac{12\lambda e^{2N\eta_0}}{\eta_0 x+4\lambda(e^{2N\eta_0}-1)}\right].
\end{eqnarray}
By imposing $P^{1/2}_R \simeq 5 \times 10^{-5}$ and $n_s=0.96$, we
plot $\phi^2/M^2_P$ (at $N=60$) and $\lambda$ as functions of
$\eta_0$ in Fig. (\ref{result}). The reason why we plot
$\phi^2/M^2_P$ is to justify that we can ignore the last term
in Eq. (\ref{potential2}).  By comparing the third and fourth terms
in Eq. (\ref{potential2}), $\lambda_p \simeq 1$ is the requirement for
 $\phi^2/M^2_P \ll \lambda$.

As we can see from Fig.~\ref{result}, when $\lambda \simeq 5\times
10^{-15}$, $\phi^2/M_P^2 \simeq 5 \times 10^{-17}$, which means the
$\phi^6$ term is at least $100$
times smaller than the $\phi^4$
term. Therefore, the last term can only
contribute about $10\%$ to $V'$
(compare $V'$ with $\phi^6$ to the original $V'$ obtained from CMB
normalization), which results in less than $10\%$ correction to $V_0$
(from CMB normalization). The contribution of the $\phi^6$ term to
$V^{\prime\prime}$ is only about $1\%$ (compare the original
$V^{\prime\prime}$ from $\eta=-0.02$ with the value coming from
the $\phi^6$ term). Nevertheless, we have a correction
about $10\%$ to $\eta$ because of the change of $V_0$,
which only affects the spectral index at the
level of $\Delta n_s\sim0.001$. Thus, we still achieve $n_s \simeq
0.96$ even if we include the $\phi^6$ term in the potential.
All these justify why we ignore the last in Eq.~(\ref{potential}) from
the beginning.

The gauge hierarchy problem requires the soft SUSY breaking parameters
in the order of $TeV$, otherwise some level of fine-tuning still exists.
The soft parameters involved in this study are $m$ and $A$, which we have
set to be $O({\rm TeV})$.  With these soft parameters we have achieved
a desirable spectral index.  As long as $m$ and $A$ are in TeV scale the
inflation with $N=60$-fold can be obtained with reasonable $\phi$ and
$\lambda$.

Since we are considering soft mass parameters in TeV scale,
supersymmetric partners of SM fermions can be readily produced at the
LHC. Once these soft mass parameters are determined, parameters for
the inflaton field of the inflationary model can be constrained.  We
will pursue this possibility in future work.

\section{Conclusions}
\label{5}

In this paper, we have shown that the potential of supernatural inflation
can be converted into a hilltop form by introducing a $A$-term (and a
negligible non-renarmalizable term). The natural value for the soft
SUSY breaking parameters are of the order TeV, which is exactly
the order that we need to reduce the spectral index,
without fine-tuning, into the latest result according to WMAP.

From Eq. (\ref{potential2}), we are assuming that $V_0$ is at the scale of gravity mediated SUSY breaking (and for some versions of gauge mediation). But the SUSY breaking scale in the framework of gauge mediation can be right down to $10^3 \mbox{ GeV}$ \cite{Giudice:1998bp}
. With the hilltop form of our potential (but \emph{not} with the original supernatural potential keeping only the quadratic term), we can accommodate these low scales. It is interesting to notice that even $V_0=0$, the model can still work (it is the MSSM inflation model). In that case we have a non-hybrid model. With lower $V_0$ there may be a model between hybrid and non-hybrid. We will consider these issues in future work.

\section*{Acknowledgement}
This work was supported in part by the
NSC under grant No. NSC 96-2628-M-007-002-MY3, by the NCTS, and by the
Boost Program of NTHU.

We are grateful to D. H. Lyth and J. McDonald for insightful comments.

\end{document}